# High-Resolution Altitude Profiles of the Atmospheric Turbulence with PML at the Sutherland Observatory

L. Catala[1,2]⋆, A. Ziad[3], Y. Fanteï-Caujolle[3], S.M. Crawford[1], D.A.H. Buckley[4,1], J. Borgnino[3], F. Blary[3], M. Nickola[5] and T. Pickering[6]

[1]*South African Astronomical Observatory, Observatory Road, Observatory 7935, South Africa*
[2]*University of Cape Town, Private Bag X3, Rondebosch 7701, South Africa*
[3]*Université Côte d'Azur, OCA, CNRS, Laboratoire J.L. Lagrange UMR 7293, Parc Valrose F-06108 Nice Cedex 2, France*
[4]*Southern African Large Telescope, P.O. Box 9,Observatory 7935, South Africa*
[5]*Space Geodesy Program Hartebeesthoeck Radio Astronomy Observatory, PO Box 443, Krugersdorp 1740, South Africa*
[6]*Space Telescope Science Institute, 3700 San Martin Drive, Baltimore, MD 21218, USA*

**ABSTRACT**
With the prospect of the next generation of ground-based telescopes, the extremely large telescopes (ELTs), increasingly complex and demanding adaptive optics (AO) systems are needed. This is to compensate for image distortion caused by atmospheric turbulence and fully take advantage of mirrors with diameters of 30 to 40 m. This requires a more precise characterization of the turbulence. The PML (Profiler of Moon Limb) was developed within this context. The PML aims to provide high-resolution altitude profiles of the turbulence using differential measurements of the Moon limb position to calculate the transverse spatio-angular covariance of the Angle of Arrival fluctuations. The covariance of differential image motion for different separation angles is sensitive to the altitude distribution of the seeing. The use of the continuous Moon limb provides a large number of separation angles allowing for the high-resolution altitude of the profiles. The method is presented and tested with simulated data. Moreover a PML instrument was deployed at the Sutherland Observatory in South Africa in August 2011. We present here the results of this measurement campaign.

## 1 INTRODUCTION

Over the past decades a number of instruments have been developed in order to measure the atmospheric turbulence, which affect the quality of images from ground based optical telescopes.

The differential image motion monitor (DIMM, Sarazin & Roddier (1990)) and the multi-aperture scintillation sensor (MASS, Kornilov et al. (2002)) are the most commonly used instruments for continuous monitoring at observatories around the world. Other instruments, such as the generalized seeing monitor (GSM, Martin et al. (1994)), the slope detection and ranging (SLODAR, Wilson (2002)) and the scintillation detection and ranging (SCIDAR, Fuchs et al. (1998)), have been extensively used during site testing campaigns. Those instruments can be classified in 2 main categories: the instruments that only measure atmospheric turbulence parameters values integrated through the entire atmosphere and the profilers providing an estimations of the turbulence profile via the turbulence structure function ($C_n^2(h)$), which gives a measure of the turbulence strength of a layer at an altitude $h$. However, all profilers have limitations: either a low-resolution altitude profile of the whole atmosphere or a high-resolution altitude profile of only a section of the atmosphere at ground layer(GL) or in the free atmosphere(FA). Despite their limitations these instruments have provided very useful information for site selection, continuous seeing monitoring, and the determination of the essential parameters needed for the design of adaptive optics (AO) systems.

With the advent of the extremely large telescopes (ELTs) with diameters greater than 30 m, the constraints on the design of an AO system are becoming more demanding, requiring the development of new atmospheric turbulence monitoring instruments that provide more accurate measures of atmospheric profiles with high resolution through the entire atmosphere. The quality of the AO correction over a large field of view for this next generation of ground-based telescopes relies on the accurate determination of the optical parameters of the atmospheric turbulence (Costille & Fusco

⋆ E-mail:lcc@saao.ac.za

2011). In particular, the distribution of the turbulence in the different layers of the atmosphere is a critical parameter.

A new instrument, Profiler of Moon Limb (PML, Ziad et al. (2013)), has been developed in order to provide $C_n^2(h)$ profiles from differential measurement of the wavefront Angle of Arrival (AA, Borgnino (1990)) fluctuations along the lunar limb. The differential measurement is made possible by the use of 2 sub-apertures, similar to the DIMM technique. Direct measurement methods use series of single images that are affected by telescope vibrations and wind shake. We get rid of these effects thanks to the differential method. Measurements are done from the difference between the measured Moon edge position in one sub-aperture with that measured in the other. Since both apertures are similarly affected by telescope vibrations and wind shake, those are suppressed by the differential measurement. Moreover, the use of the Moon limb offers the advantage of providing a very high-resolution altitude profile of the turbulence ($C_n^2(h)$) in addition to all the integrated atmospheric parameters that other instruments provide: the coherence length ($r_0$), the seeing ($\varepsilon_0$), the coherence time ($\tau_0$) and the isoplanatic angle ($\theta_0$). The PML instrument is an expansion of the monitor of outer scale (MOSP, Maire et al. (2007)) concept based on a direct measurement method using series of single images of the Moon limb to retrieve the outer scale profile ($\mathcal{L}_0(h)$) from the structure function of transverse AA fluctuations. Similarly, the PML provides a measure of the turbulence profile ($C_n^2(h)$) from the differential covariance of these fluctuations.

In the context of the Sutherland site characterization and in order to provide information for a simulation study on potential AO performances on the Southern African Large Telescope (SALT), a PML observing campaign was carried out at Sutherland. This campaign was primarily used to work on the instrument data processing and inversion method development, but also provided first results from PML measurements at the Sutherland site.

In this paper we present the theoretical background of the PML working principle in the second section. An overview of the optical layout of the instrument is given in section 3. The fourth section is dedicated to the measurement and data processing technique.

We used simulated data to probe the validity of the method and test our data analysis process. Those simulations are presented in the fifth section. In August 2011 the PML was deployed at the Sutherland SAAO Observatory in South Africa. The results of this observing campaign, as well as comparison with ancillary instruments, are presented in the sixth section. In the seventh section we proffer concluding remarks about the PML method.

## 2 THEORETICAL BACKGROUND AND RECONSTRUCTION METHOD

The PML uses Moon images from two sub-apertures in order to measure the profile of the atmospheric turbulence as illustrated in Figure 1. In order to recover the turbulence profile ($C_n^2(h)$) we compare our data to a theoretical model. The method uses the covariance of the transverse AA fluctuations. Comparing the theoretical and measured covariance is a non-linear inverse problem that, for the Sutherland

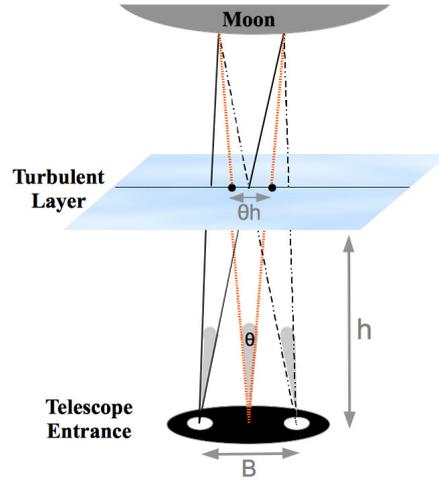

**Figure 1.** Principle of PML measurement. When measuring the angular covariance of a system with a fixed base, $B$, the contribution of a layer at an altitude $h$ peaks for an angular value of $\theta = \frac{B}{h}$.

campaign, we chose to resolve using a simulated annealing method (Maire et al. 2007).

Here we recall the theoretical expression of the spatio-angular covariance of the transverse AA fluctuations in the case of the Von Karman turbulence model (Von Karman 1948), detail how we compute it, and test the response of our inversion grid.

### 2.1 Theoretical Angle of Arrival covariance

The optical atmospheric turbulence is commonly described by the spectrum of its index of refraction ($n$) fluctuations that follows a Kolmogorov's law: $\Phi_n(k) = 0.033\ C_n^2\ k^{-11/3}$, where $k$ is the wave-number. However, the Kolmogorov's model assumes an infinite outer scale ($\mathcal{L}_0$) value. In order to take into account the finite size of the outer scale, other models were developed. In the case of the PML instrument we use the Von Karman model: $\Phi_n(k) = 0.033\ [2\pi]^3\ C_n^2\ [k^2 + [\frac{2\pi}{\mathcal{L}_0}]^2]^{-11/6}$. Based on this model, the AA spatial covariance is given by (Borgnino et al. 1992; Avila et al. 1997):

$$C_\alpha(B, D) = 1.19 \sec(z) \int dh C_n^2(h) S(B, D, \mathcal{L}_0(h)), \quad (1)$$

with

$$S(B, D, \mathcal{L}_0(h)) = \int df f^3 (f^2 + \frac{1}{\mathcal{L}_0(h)^2})^{-11/6}$$

$$[J_0(2\pi f B) + J_2(2\pi f B)] \left[ 2 \frac{J_1(\pi D f)}{\pi D f} \right]^2, \quad (2)$$

where $z$ is the zenith angle, $B$ is the separation between two sub-apertures of diameter $D$, $\mathcal{L}_0(h)$ is the wavefront outer scale at the altitude $h$, $f$ the spatial frequency and $J_m$ are Bessel function of order $m$.

In the case of differential measurements, and for observations in two directions separated by an angle $\theta$ (Figure 1) the differential angular covariance can be expressed as follows (Ziad et al. 2013):

$$C_{\Delta\alpha}(B, D, \theta) = 2C_\alpha(\theta h, D) - C_\alpha(B - \theta h, D) - C_\alpha(B + \theta h, D). \quad (3)$$

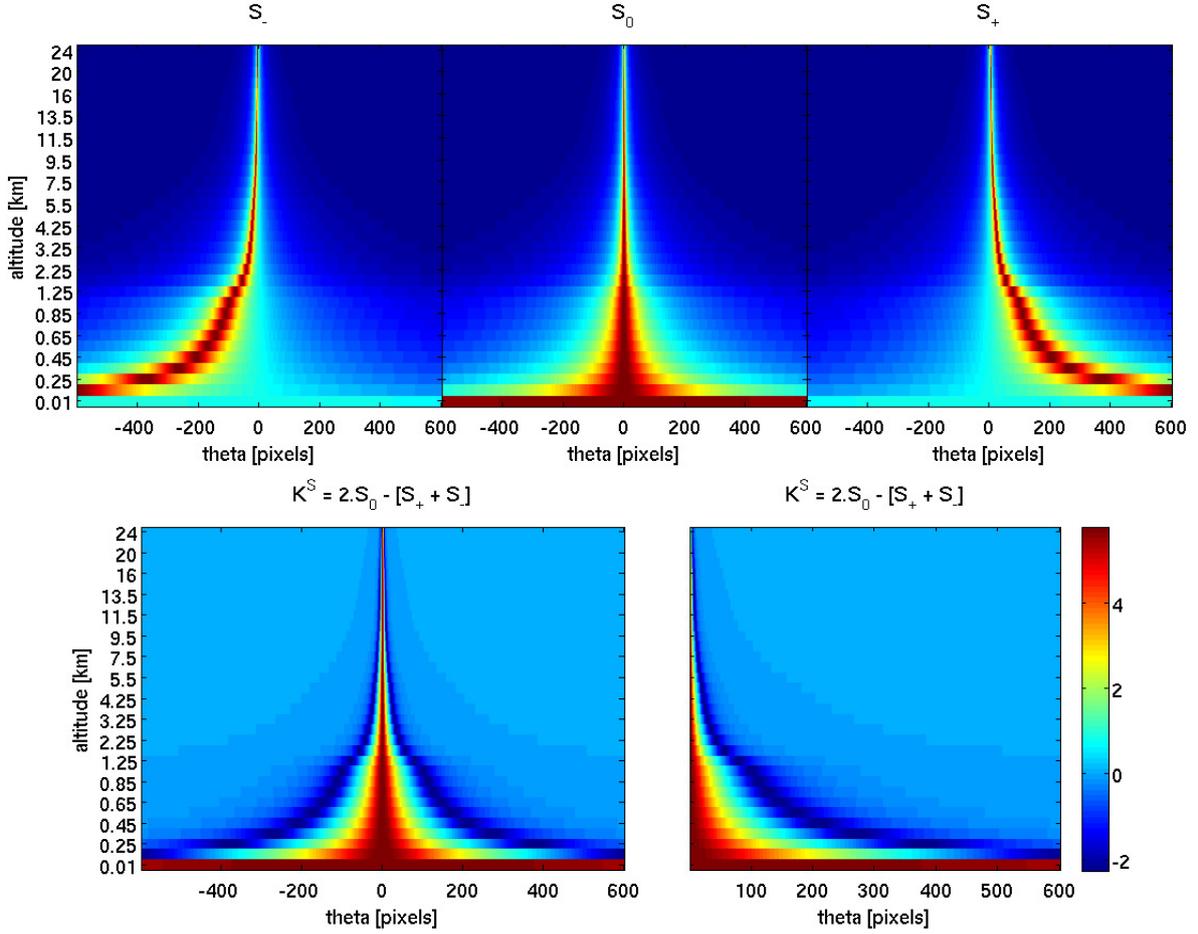

**Figure 2.** Theoretical "S" functions, given by eq. 6. Top: $S_-(\theta, h)$, $S_0(\theta, h)$ and $S_+(\theta, h)$ from left to right. Bottom: $K^S$ matrix. In all five figures h is increasing from bottom to top. Values are given for the 33 single layers of the reconstruction grid. $\theta$ goes from -356 to 356 arcseconds from left to right for the three top figures and the bottom left. The bottom right figure shows the positive values of $K^S$ that we will use for the inversion, as we only measure positive $\theta$.

$\theta h$ is the spatial distance of the perturbed wavefront intercepted by an angle $\theta$ at an altitude $h$ (Figure 1).
Using eq. 1 and 2, this gives:

$$C_{\Delta\alpha}(B, D, \theta) = 1.19 sec(z) \int dh C_n^2(h)[2S_0^h - S_-^h - S_+^h], \quad (4)$$

where, $S_0^h = S(\theta h, D, \mathcal{L}_0(h))$, $S_-^h = S(B - \theta h, D, \mathcal{L}_0(h))$ and $S_+^h = S(B + \theta h, D, \mathcal{L}_0(h))$.
Considering the overall atmosphere as a superposition of thin $\Delta h_i$ discrete layers at altitudes $h_i$ we can rewrite this expression as a sum:

$$C_{\Delta\alpha}(B, D, \theta) = 1.19 sec(z) \sum_i \Delta h_i C_n^2(h_i)[2S_0^{h_i} - S_-^{h_i} - S_+^{h_i}], \quad (5)$$

For easier calculation, we split the components solely dependent on predefined parameters (altitude grid and system parameters) from those dependent on parameters that need to be determined($C_n^2(h)$, $\mathcal{L}_0(h)$), as follows:
- The energy term, containing the turbulence strength information:

$K^{Cn}(h) = 1.19 sec(z) \Delta h C_n^2(h)$.
- The shape term, containing the outer scale value information:
$K^L(h, f) = f^3(f^2 + \frac{1}{\mathcal{L}_0(h)^2})^{-11/6}$.
- The filtering terms, linked to the system sub-pupils and base:
$K_0^J(h, \theta, f) = [J_0(2\pi f \theta h) + J_2(2\pi f \theta h)] \left[ 2\frac{J_1(\pi D f)}{\pi D f} \right]^2$,
$K_-^J(h, \theta, f) = [J_0(2\pi f (B - \theta h)) + J_2(2\pi f (B - \theta h))] \left[ 2\frac{J_1(\pi D f)}{\pi D f} \right]^2$
and
$K_+^J(h, \theta, f) = [J_0(2\pi f (B + \theta h)) + J_2(2\pi f (B + \theta h))] \left[ 2\frac{J_1(\pi D f)}{\pi D f} \right]^2$.
This allows us to rewrite the S integrals in the following form:

$$S_{0,-,+}(h, \theta) = \int_f df K^L(h, f) . K_{0,-,+}^J(h, \theta, f), \quad (6)$$

The $S_{0,-,+}(h, \theta)$ functions can be determined for each indi-

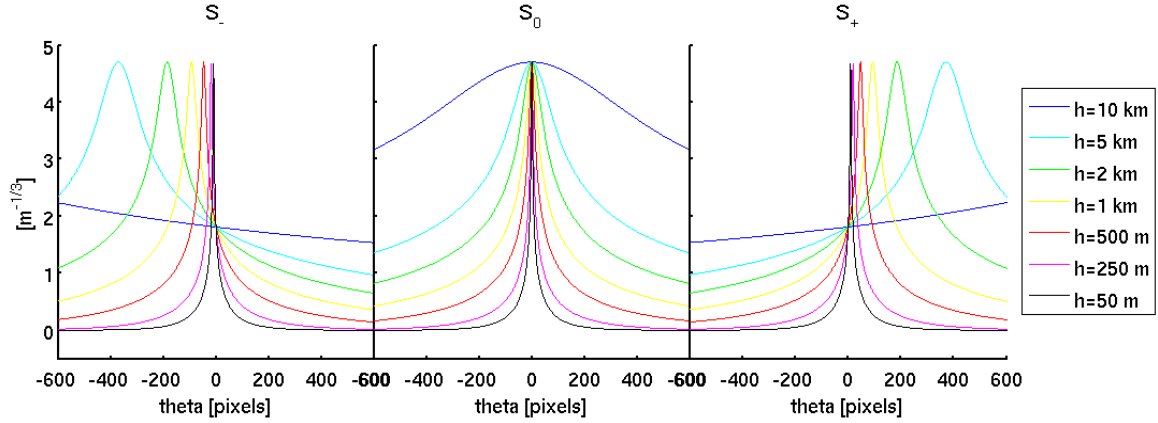

**Figure 3.** Theoretical $S_{0,-,+}$ for single layers with $\mathcal{L}_0(h) = 20m$. Left: $S_-$. Center: $S_0$. Right: $S_+$. Each color represents the functions for a different layer altitude: h= 50m, 250m, 500m, 1km, 2km, 5km and 10km. Each lines on these figures correspond to single rows from the top three images of Figure 2.

vidual layer and hence, summing over all altitudes, gives:

$$C_{\Delta\alpha}(\theta) = \sum_h K^{Cn}(h).[2\int_f df K_0^J(h,\theta,f).K^L(h,f)$$
$$- \int_f df K_+^J(h,\theta,f).K^L(h,f) - \int_f df K_-^J(h,\theta,f).K^L(h,f)]. \quad (7)$$

If we consider the case of a fixed $\mathcal{L}_0$, the three $\int_f df K^J.K^L$ components can be precalculated and stored in a matrix $K^S = 2S_0 - S_- - S_+$ (Figure 2). We can then write:

$C = K^{Cn}.K^S,$

where $K^{Cn}$ is a 1xN matrix and $K^S$ is a NxM matrix, with N the number of layers of the reconstruction grid and M the number of separation angles ($\theta$) along the Lunar limb.

Using a chosen altitude grid and the $\theta$ values set by the system configuration, we can compute all the $K^J$ functions and, in turn, the $S$ functions at fixed $\mathcal{L}_0$ and the corresponding $K^S$ matrix. The top row of Figure 2 shows the theoretical $S_{0,-,+}$ functions while the bottom graphs represent the $K^S$ matrix that will be used for the inversion. The bottom left shows the full $K^S$ matrix, including negative $\theta$ values. As we only perform measurements for positive $\theta$, we will use the positive side of the matrix for the inversion shown on the bottom right side figure. The number of separation angles available (x-axis) is determined by the system layout and is given by the number of pixels along the Lunar limb. Here we have set the number of layers to 33 with a range of altitudes going from 10 m to 24 km above the telescope entrance pupil. $\mathcal{L}_0$ is set to 20 m. We also show the 2D curve of the theoretical $S_{0,-,+}$ functions for 7 individual layers in Figure 3. From this representation, we can clearly see that the position of the peak of covariance in the lateral components (left and right figures) is dependent on the layer altitude $h$: $\theta_{peak} = \frac{B}{h}$, with B the base between the 2 sub-apertures at the telescope entrance pupil. Hence, for lower altitudes the peak of covariance is located at larger separation angles $\theta$.

Figure 4 shows the combined $S_0, S_-$ and $S_+$ (blue lines)

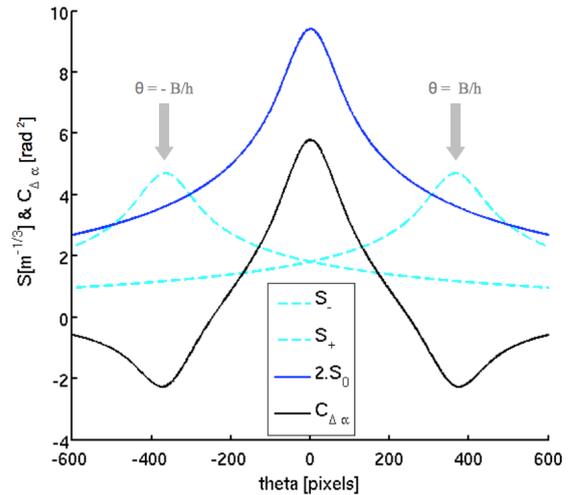

**Figure 4.** Covariance for a single layer of unity strength ($K^{Cn} = 1$) at h=350 m and with $\mathcal{L}_0 = 20m$. Blue line: $2S_0$. Blue dotted lines: $S_{+,-}$. Note the position of the peak of the lateral component located at $+/-\frac{B}{h}$. Black line: $C_{\Delta\alpha}(h=1km) = K^{Cn}.[2S_0-S_--S_+]$. Note the high value of $C_{\Delta\alpha}$ due to the fact that we took $K^{Cn} = 1$ for simplicity and better visualization of all curves together.

resulting in the theoretical differential covariance (black line) for a single layer at 350m. Here, for simplicity we consider a layer of unity strength, $K^{Cn} = 1$, and $\mathcal{L}_0 = 20m$. Note that we only measure positive separation angles, therefore, for the reconstruction we only consider the positive components of the covariance.

In the more realistic case of an atmosphere made up of multiple layers of variable thickness at different altitudes, with different turbulence strength and outer scale value, one will input each of the layers parameters ($h_i$, $\Delta h_i$, $C_n^2(h_i)$ and $\mathcal{L}_0(h_i)$), before adding up all the layers contribution to get the equivalent covariance for the overall atmosphere. From those parameters, we chose two beforehand ($h_i$, $\Delta h_i$). The other two ($\mathcal{L}_0(h_i)$, $C_n^2(h_i)$), can be retrieved by minimizing the difference between theoretical and measured values. However, while the different components ($S_0^h, S_-^h, S_+^h$) of the differential angular covariance have a strong dependence on the outer scale, its impact on the differential covariance itself is mitigated by the fact that it is given by the combination of twice the central covariance minus the two lateral covariance. It is only when the outer scale is small (in the metric range) that its impact on the differential covariance cannot be neglected anymore (Borgnino et al. 1992). In the particular case of astronomical observatory sites the outer scale is known to be in the decametric range. Hence we can simplify the inversion problem by taking a fixed the outer scale value and reduce the reconstruction to the turbulence profile alone. Similarly to the work done on the MOSP instrument (Maire et al. 2007), we will use a simulated annealing algorithm for the minimization process leading to the reconstruction of the turbulence profile, $C_n^2(h).dh$. Other minimization technique were also tested and presented in (Blary et al. 2014).

Table 1. Reconstruction altitude grid.

| GL | | | | h[m] | 10 | 150 | 250 | 350 | 450 | 550 | 650 | 750 | 850 | 950 | | | | | |
|---|---|---|---|---|---|---|---|---|---|---|---|---|---|---|---|---|---|---|---|
| | | | | | | | | | dh = 100m | | | | | | | | | | |
| FA | | | | h[km] | 1.25 | 1.75 | 2.25 | 2.75 | 3.25 | 3.75 | 4.25 | 4.75 | | | | | | | |
| | | | | | | | | | dh = 500m | | | | | | | | | | |
| h[km] | 5.5 | 6.5 | 7.5 | 8.5 | 9.5 | 10.5 | 11.5 | 12.5 | 13.5 | 14.5 | | | 16 | 18 | 20 | 22 | 24 | | |
| | | | | | | dh = 1 km | | | | | | | | | dh = 2 km | | | | |

### 2.2 Altitude grid and Inversion Response

In order to cover both the GL and the FA part of the atmospheric turbulence, we chose a 33 layers grid. There are 10 layers for the GL below 1 km and 23 layers for the FA between 1 and 25 km. The detail of the altitude grid is given in Table 1. The number of layers was chosen in agreement with previous findings that showed the necessity to know 30 to 40 layers in order to feed the later adaptive optics systems using tomography (Costille & Fusco 2012).

We tested the response of the reconstruction grid to 89 individual turbulent layers with altitudes ranging from 5 m to 30 km. For simplicity all layers were of unit strength ($K^{Cn}$ = 1) and with an outer scale value ($\mathcal{L}_0$) fixed to 20 m. We show the response results in Figure 5. In the top figure, the x-axis shows the altitudes of the input turbulent layer, while the y-axis shows the altitudes of the reconstruction grid. The pink ellipses represent the relative amount of turbulence in each layer of the reconstruction grid. The bottom graph of Figure 5 shows how each of the reconstruction grid altitude is sensitive to turbulence in the adjacent layers. The base of each triangle gives the range of altitudes for which the individual layers of the reconstruction grid can partially sense turbulence. The height at each altitude gives the sensitivity strength from 0 to 1, the latter being 100% sensitive.

For each of the 89 single turbulent layers the reconstruction process should apportion the turbulence of the input layer between the 33 layers of the reconstruction grid. One expects that for a turbulent layer located at one of the reconstruction grid altitudes, all the turbulence will be reflected in that layer after the inversion. In the case of a turbulent layer located in between two altitudes of the reconstruction grid, one will expect the reconstruction to spread the turbulence between the adjacent layers. If we take the input 10 km layer (on the x-axis), located between the 9.5 km and 11 km layers of the reconstruction grid (on the x-axis), the turbulence is redistributed with 63.7% in the 9.5 km layer and 36.3% in the 11 km one. Similarly the 21 km layer is split with 48.15% in the 20 km layer and 51.85% in the 22 km layer. The redistribution agrees with the theoretical expectations and validates both the choice of our altitude grid and inversion method.

## 3 OPTICAL LAYOUT AND MEASUREMENT METHOD

### 3.1 PML optical layout

The PML (Ziad et al. 2010, 2013) was designed to provide high-resolution altitude profiles of the atmospheric turbulence. Similar to the DIMM technique, it uses a differential method via a two sub-aperture mask mounted at the entrance pupil of the telescope, allowing telescope vibration and wind shake effects to be ignored. The profiles are reconstructed from the differential covariance functions. The use of the continuous Lunar limb, as compared to a double star with SLODAR, provides a large number of separation angles, allowing for the high-resolution of the altitude profiles.

The PML consists of a 16-inch MEADE telescope tube mounted on an Astro-Physics AP3600 equatorial mount with a mask made of 2 holes with separation $B = 0.267m$, and diameter $D = 0.06m$ (Figure 6, left). When pointing the telescope at the Moon two images of the limb are produced, corresponding to the two sub-apertures. In order to separate the 2 images, a Dove prism (D) is introduced in the optical path (Figure 6, right). The Dove prism flips over one of the images and avoids overlap of the images. The image acquisition is performed by a PCO Pixelfly CCD operating at a frame rate of 33 Hz. The CCD, with a pixel size of 9.9 microns, produces images of 640x480 pixels. The image scale is 0.594 arcsec per pixel. The exposure time needs to be short enough to "freeze" the turbulence, typically of the order of a few ms (i.e. $\tau_0$). Here it was set to 5 ms. The number of images used for each measurement was set to a thousand images per data set. For each acquisition we used the statistical properties of the atmospheric turbulence to retrieve its parameters.

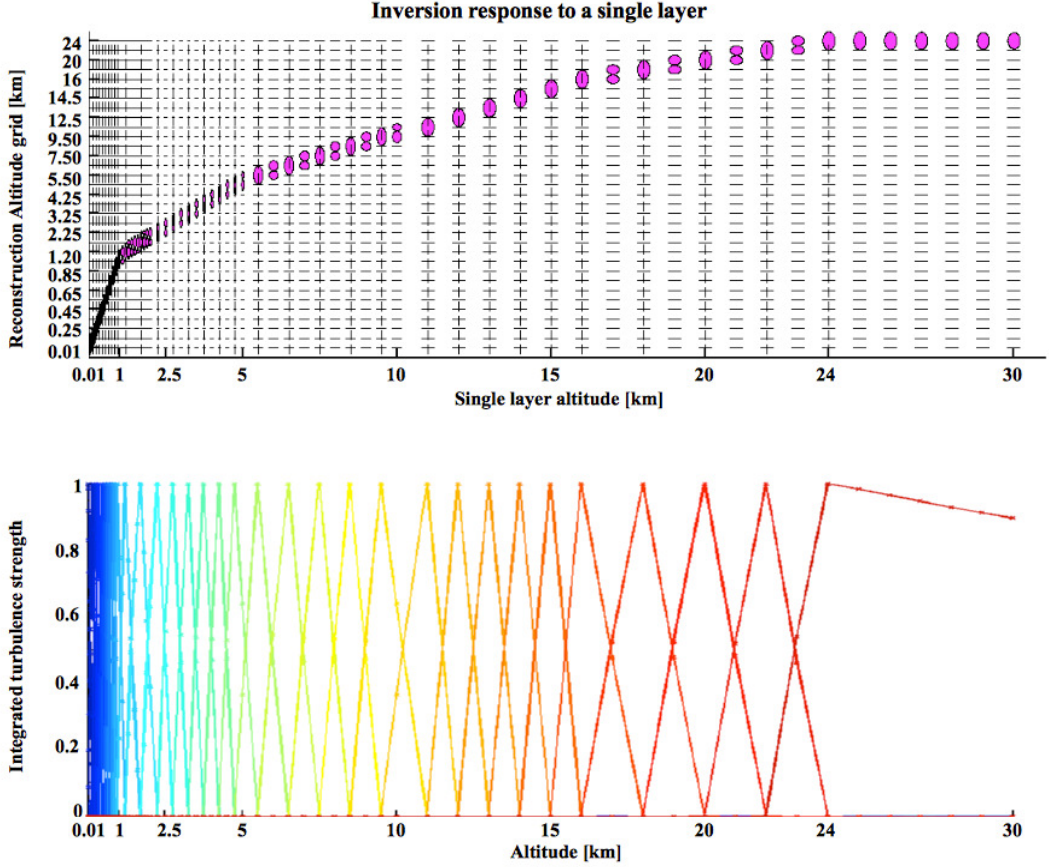

**Figure 5.** Response of the inversion algorithm to single turbulent layers. Top: the single turbulent layer input altitude is given on the x-axis while the y-axis represents the altitudes of the reconstruction grid. For each input layer (x) the relative distribution of the turbulence throughout the reconstruction grid layers (y) is represented by the pink ellipses. Bottom: The color triangle-like shaped curves represent the sensitivity of each altitude grid height to turbulence in all 89 input layers.

### 3.2 Image pre-processing and "cleaning"

Prior to the data analysis that will lead to the profile reconstruction, there are a number of steps that need to be followed to make sure that we removed any instrumental bias due to optical misalignment and imperfect tracking. This will ensure that when performing the differential measurements we properly match the same point on the Moon edge from both images, hence only measuring the edge motion due to atmospheric turbulence. The full pre-processing, summarized in Figure 7, involves the following:
1. Flat fielding and Dark frame subtraction.
2. Measuring image rotation, should any remain after the optical alignment of the Dove prism.
3. Measuring shift in the x-direction, if any.
4. Measuring image drift due to telescope pointing inaccuracy. if any.
5. Applying, rotation, shift and drift correction to the images.

The eventual residual rotation is measured by stacking all images of an acquisition. This is equivalent to a long-exposure image and suppresses the seeing effect to only keep the static optical misalignment. The difference between the top and bottom edge positions gives us the residual rotation angle. The x-drift is measured for each image as compared to a reference image chosen to be the first of an acquisition. Finally the drift parallel to the y-axis due to imperfect polar alignment is measured by fitting a line to the data representing the mean edge position throughout an acquisition as shown in the image 4 of Figure 7.

After applying all corrections, we can measure the position of the Moon edges on the images that we will use for the profile reconstruction.

### 3.3 Angle of arrival covariance - Experimental measurements and profile reconstruction

For each image, we determine the edge position by taking the image derivative before using a barycenter method for the detection (Maire et al. 2007). The method is illustrated in Figure 8. Once the edge positions have been determined for all sets of two images of an acquisiton, we can calculate the experimental differential covariance of the AA as illustrated

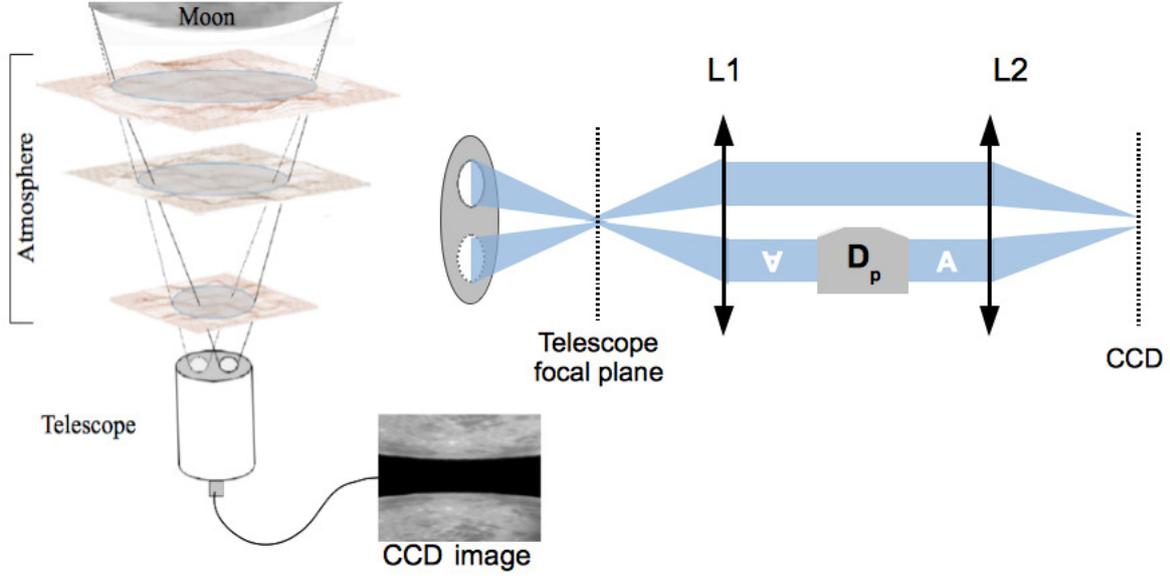

**Figure 6.** PML optical layout. Left: Schematic of the overall instrument setup with the 2 sub-aperture mask at the entrance pupil of the telescope. Right: Schematic of the optical path from the telescope entrance pupil to the imaging CCD. L1 is a collimating lens, $D_P$ represents the Dove prism and L2 refocusses the collimated beam onto the imaging CCD.

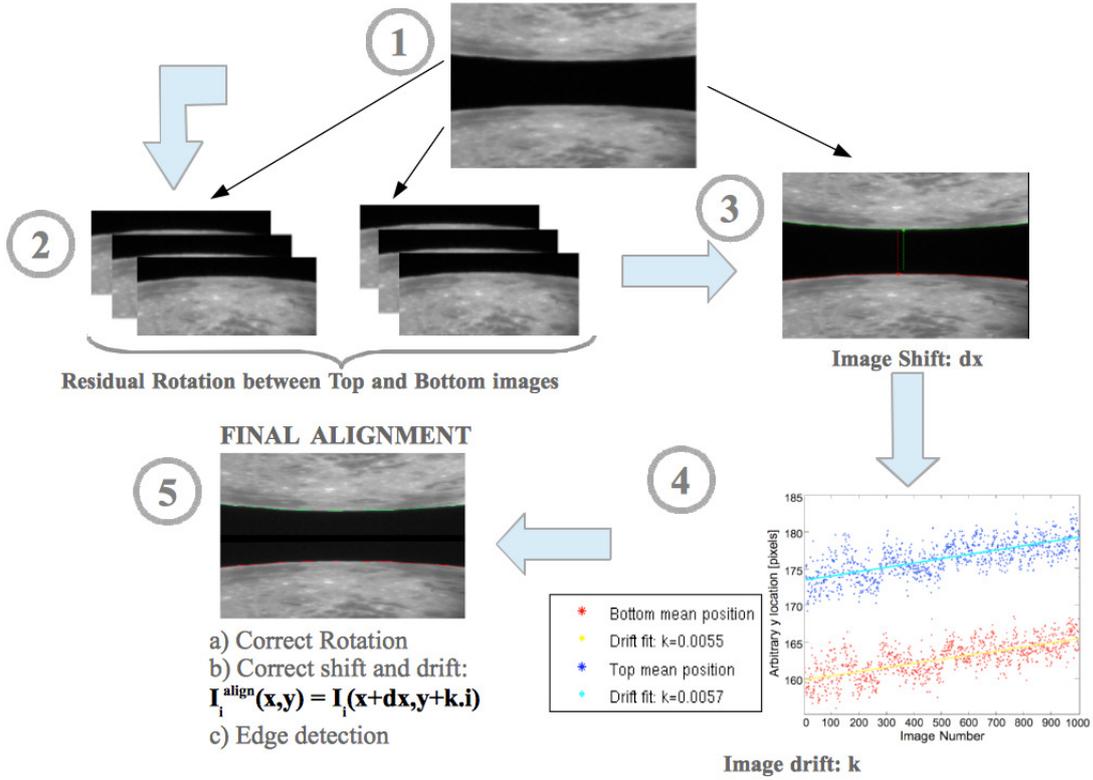

**Figure 7.** Full alignment process summary. 1: Flat fielding and Dark frame substraction. 2: Measure the residual rotation between the Top and Bottom images. 3: Find the x-displacement of each image ($dx_i$) with respect to the reference image. 4: Find the amount of drifting between images (k). The figure in step 4 shows the averaged position of the limb (y) for all 1000 images of an acquisition (x) and the slope of line fits (cyan and yellow) gives us the amount of drifting between two consecutive images (dark blue dots and cyan line are from the Top image while red dots and yellow line are from the Bottom image). 5: Applying the shift ($dx_i$) and drift ($k.i$) correction to each image, after correction for the rotation. Then the images are ready to be used for edge detection and data extraction.

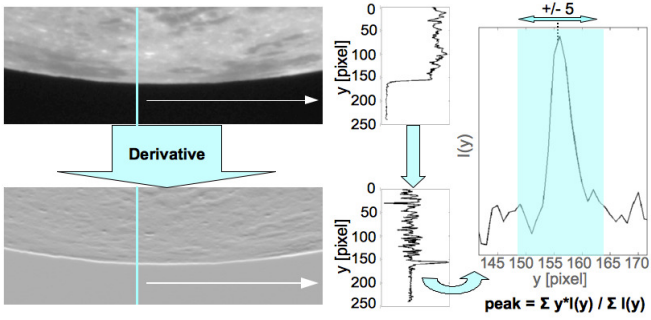

**Figure 8.** Edge Detection. Top left: original image and an example of a vertical cut showing a Heavyside step function at the edge of the Lunar limb. Bottom Left: Derivative of the image and the peak function of its vertical cut. Right: Zoom in around the peak of the image derivative and windowing used to perform a barycenter measurement of the peak position.

in Figure 9:

$$C_{\Delta_\alpha}^{meas}(\theta) = \langle [\alpha_T(x_i) - \alpha_B(x_i)][\alpha_T(x_i + \theta) - \alpha_B(x_i + \theta)] \rangle,$$

where $\alpha_T(x_i)$ and $\alpha_B(x_i)$ are the vertical coordinates of the limb at an initial $x_i$ coordinate for, respectively, the top and bottom images. $\alpha_T(x_i + \theta)$ and $\alpha_B(x_i + \theta)$ are the positions at the $x_i$ coordinate along the limb separated by an angle $\theta$ from the initial position. The brackets signs ($\langle \rangle$) represent the average value for all the products corresponding to a specific separation angle $\theta$ along the edge. One can see that the larger $\theta$, the fewer number of measurements along the finite length of the Lunar limb. After the image pre-processing, the "cleaned" images are generally between 550 and 600 pixels wide. This gives a maximum of 599 measurements for the smallest separation angle ($\theta_1 = 1$ pixel $= 0.594$ arcseconds) and a single measurement for the largest separation angle ($\theta_{max} \sim 599$ pixels $\simeq 356$ arcseconds). The measurement error is therefore much larger for larger than smaller $\theta$. In the inversion process we will weigh the fits by the number of data points for each $\theta$. For one acquisition, we calculate $C_{\Delta_\alpha}(\theta)$ for each of the 1000 images. The final differential covariance function for the acquisition is obtained by taking the average of all thousands $C_{\Delta_\alpha}(\theta)$.

Using both, measured and theoretical, covariances one can reconstruct the turbulence. We generate an atmospheric profile ($h_i$, $\Delta h_i$, $C_n^2(h_i)$ and $\mathcal{L}_0(h_i)$) with which we compute the corresponding theoretical covariance function before comparing it to the measured one.

We use a simulated annealing (SA) algorithm (Kirkpatrick et al. 1983) to find the best fit value. The SA algorithm is a random search technique, which exploits an analogy with thermodynamics and the way in which a metal cools and freezes into a minimum energy, assimilated here to our global minimum. Starting from an initial set of $C_n^2(h_i)$ values we compute the initial cost function ($E_{n=0}$) between the theoretical and measured covariance: $E_{n=0} = \sum_\theta [C_{\Delta_\alpha}^{theo}(\theta) - C_{\Delta_\alpha}^{meas}(\theta)]^2$. Then, for each subsequent iteration (n), we apply a small variation to the previous $C_n^2(h_i)$ values, calculate the new cost function, and then compute the cost difference $\Delta E = E_{n+1} - E_n$. If it is negative, the cost decreases and we keep the new set of parameters. If the cost increases, $\Delta E$ is positive, we do not systematically reject the solution but accept it with a probability $p = e^{(-\Delta E/T)}$. This cost-increasing acceptance probability allows for exploring the full parameter space and avoids becoming trapped in a local minimum. This acceptance probability is set by the "temperature" parameter $T$, in analogy with thermodynamics. The SA algorithm starts with a high initial temperature to explore a wide area of the parameter space and a "cooling" schedule slowly lowers the "temperature" towards the reduction of the search around the global minimum. We will stop the search, and keep the current best set of $C_n^2(h_i)$ values as our best fit result, when, at a fixed "temperature", no improvement to the cost function can be made. A similar technique was also used in Maire et al. (2007).

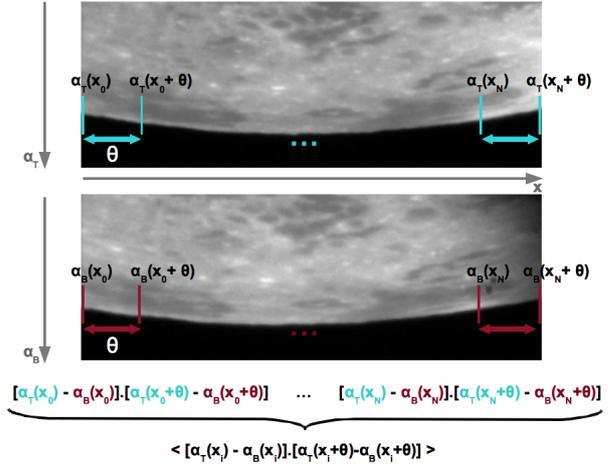

**Figure 9.** Experimental covariance. Detail of the differential covariance measurement. We calculate the product of the difference between top (blue) and bottom (red) positions at $x_i$ and $(x_i + \theta)$ for all $x_i$ positions along the edge. The average of all products is the differential covariance ($C_{\Delta_\alpha}^{meas}(\theta)$) value at a separation $\theta$

### 3.4 PML Fried parameter extraction

In addition to the turbulence profile the PML data can be used to measure the integrated seeing by determining the Fried parameter. For each acquisition, we have the temporal variation of the position over the 1000 images and for each position along the edge. This provides 600 DIMM measurements per acquisition. In the case of the PML the motion is only measured in the direction perpendicular to the Lunar limb that corresponds to the direction perpendicular to the sub-apertures separation base, hence the transverse motion. The classical relation between the Fried parameter ($r_0$) and the variance of the image position ($\sigma^2$) can be found in Fried (1966); Tatarskii (1971). In our case we will use the absolute variance ($\sigma_{abs}^2$) which is calculate from the absolute positions ($y - \langle y \rangle$) rather than the raw ones. For all positions along the edge we compute $\sigma_{abs}^2$ over the 1000 images. Then we used the expression derived by Ziad et al. (1994), based on $\sigma_{abs}^2$ and including the outer scale ($\mathcal{L}_0$):

$$r_0^{5/3} = 0.179 sec(z)\lambda^2 \frac{[D^{-1/3} - 1.52\mathcal{L}_0^{-1/3}]}{\sigma_{abs}^2}$$

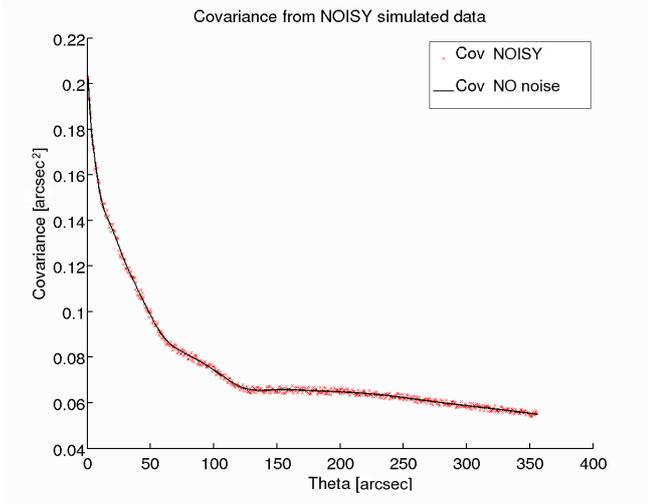

**Figure 10.** Simulated covariance for 33 layers with a profile given in the second column of Table 2 (simu). The black line shows the perfect covariance overlapped to the noisy covariance (red crosses) produced by adding Gaussian noise to the perfect curve.

We can apply this method to either the top or bottom images independently. Note that when $\mathcal{L}_0$ tends towards infinity, the expression simplifies to the more general Kolmogorov's case.

Similarly, a method that uses the differential motion have been developed for both the transverse motion, perpendicular to the direction of the sub-apertures separation, and the longitudinal motion, parallel to the direction of the sub-apertures separation. It can also be used to determine the Fried parameter. In the PML case we are only looking at the transverse differential variance ($\sigma_t^2$). The original formula was derived by Sarazin & Roddier (1990):

$$r_0^{5/3} = \lambda^2 * sec(z) * D^{-1/3} * \frac{K_t}{\sigma_t^2},$$

with,

$$K_t = 0.358 * (1 - 0.811 * S^{-1/3});$$

where $S = \frac{B}{D}$, D is the apertures diameter and B the separation between the two apertures and z is the zenith angle. An updated value of the constant $K_t$ is given in Tokovinin (2002):

$$K_t = 0.364 * (1 - 0.798 * S^{-1/3} - 0.018 * S^{-7/3}).$$

The method using top or bottom images independently was useful during preliminary tests to make sure that the results from top images were consistent with those from the bottom images. However those are strongly affected by telescope vibrations and wind shake. The measurements of the Fried parameter presented in section 6 were extracted with the more reliable differential method implemented with the later $K_t$ value.

**Table 2.** Reconstruction Profile from covariance simulations.

| Altitude | Relative strength of the turbulence layers [%] | | |
| --- | --- | --- | --- |
| | 33 simulated layers | | |
| [km] | simu | recon | recon (NOISY) |
| 0.01 | 31 | 30.6 | 31 |
| 0.15 | 6 | 6 | 6.1 |
| 0.25 | 4 | 4 | 3.8 |
| 0.35 | 3 | 2.9 | 3.3 |
| 0.45 | 2 | 2 | 1.6 |
| 0.55 | 3 | 3 | 3.4 |
| 0.65 | 2 | 2 | 1.5 |
| 0.75 | 5 | 5 | 5.5 |
| 0.85 | 5 | 5 | 4.6 |
| 0.95 | 2 | 2 | 2.3 |
| 1.2 | 2 | 2 | 1.9 |
| 1.7 | 2.5 | 2.5 | 2.5 |
| 2.25 | 2 | 2 | 2 |
| 2.75 | 1.5 | 1.5 | 1.6 |
| 3.25 | 1 | 1 | 0.6 |
| 3.75 | 5 | 5 | 6 |
| 4.25 | 3 | 3 | 1.6 |
| 4.75 | 1.6 | 1.6 | 2.9 |
| 5.5 | 1.4 | 1.4 | 0.6 |
| 6.5 | 0.5 | 0.5 | 1.1 |
| 7.5 | 1.4 | 1.4 | 0.9 |
| 8.5 | 2.2 | 2.2 | 2.7 |
| 9.5 | 2.3 | 2.3 | 2 |
| 11 | 2 | 2 | 2.2 |
| 12 | 1.5 | 1.5 | 1.6 |
| 13 | 2.1 | 2 | 1.7 |
| 14 | 0.3 | 0.3 | 0.5 |
| 15 | 0.5 | 0.5 | 0.8 |
| 16 | 0.9 | 0.9 | 0.7 |
| 18 | 0.7 | 0.7 | 0.6 |
| 20 | 0.3 | 0.9 | 0.4 |
| 22 | 1.2 | 1.2 | 1.7 |
| 24 | 1.1 | 1.1 | 0.4 |

## 4 SIMULATIONS

In order to probe our reconstruction method we simulated differential covariances for a profile with altitudes matching our reconstruction grid. We looked at two cases, one with a perfect covariance curve and one with a noisy covariance curve. The noisy data were produced from a perfect covariance to which we added Gaussian noise (Figure 10). The additional noise is within 5 per cent of the value of the "clean" simulated data.

We show the reconstruction results in Figure 11, with the input simulated data in red and reconstruction in blue. On the left we show the covariances while on the right we have the corresponding turbulence profiles. For both graphs we give the mean relative error between the input data and the output reconstruction. The top two panels correspond to the perfect covariance case, while the bottom panels correspond to the noisy data case. In addition, the relative strength of the layers from the simulated and reconstructed profiles are reported in Table 2.

The relative error of the reconstruction from a perfect data set seems negligible on the covariance, with a value close to zero. However, this still reflects as a 1.57% relative error on the reconstructed profile. In the case of the noisy data set, the relative error on the reconstructed covariance is 0.45%, which reflects as 13.5% on the reconstructed profile. The error on the profile is more important for the higher layers of the atmosphere. At higher altitudes the covariance peaks from different layers get closer to each other (Figure 3) and hence the response of the reconstruction is more sensitive to turbulence in adjacent layers, seen as wider triangles in the bottom graph of Figure 5. As a result we should see some error coming from an incorrect redistribution between adjacent layers. In order to evaluate this effect, we compared the reconstructed and original simulated profiles after applying a smoothing over three consecutive layers: $C_n^2(h_i) = \frac{1}{3} \sum_{k=i-1}^{i+1} C_n^2(h_k)$. After smoothing, the relative error between the profiles goes down to 4.3%, confirming that a large part of the error originates from an incorrect redistribution between adjacent layers. Also, in some cases, poorer optimization of the algorithm could generate convergence issues and additional error in the redistribution.

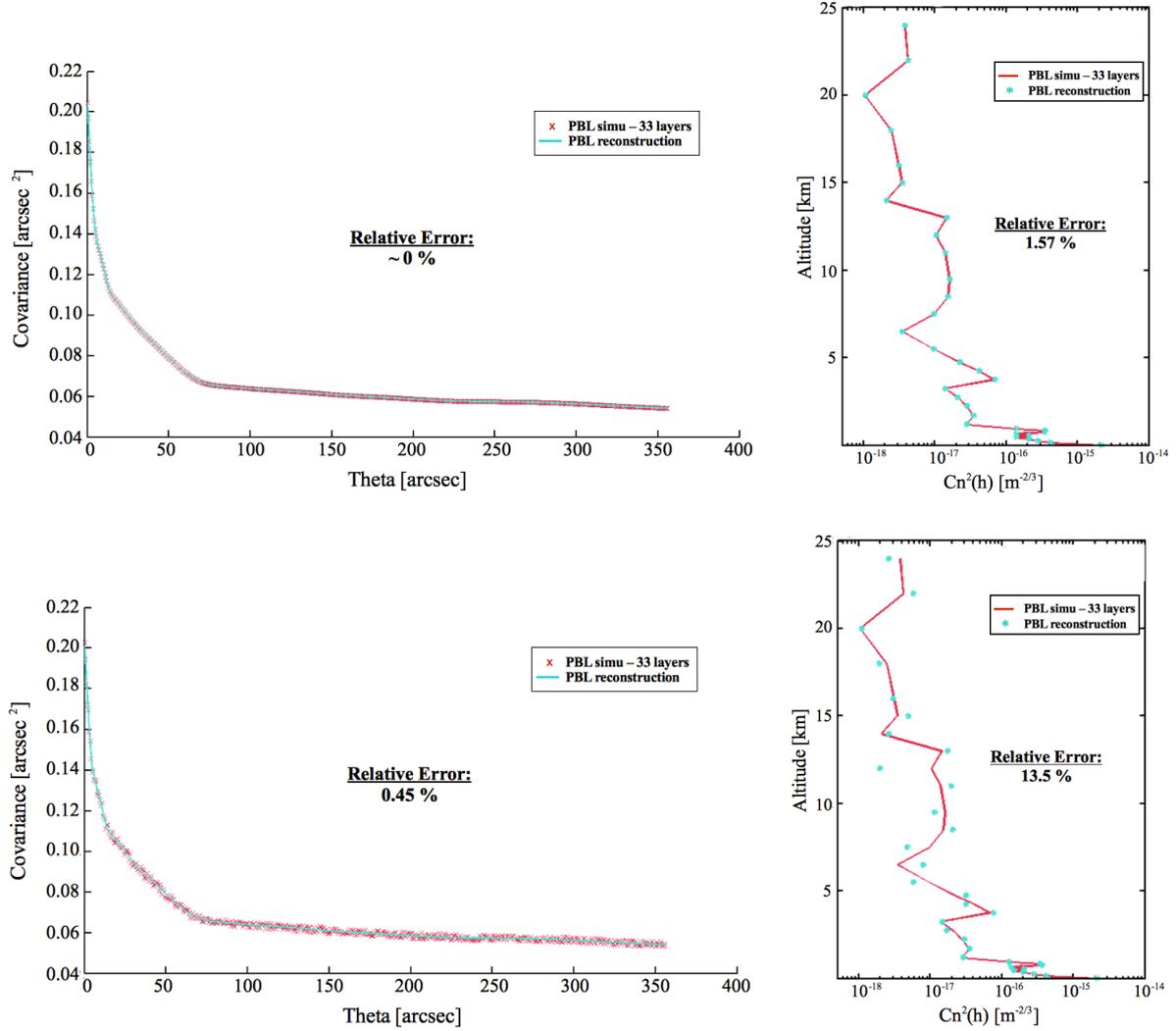

**Figure 11.** Example of simulated Covariance and the best fit from our reconstruction method in the case of a 33 layers model matching the reconstruction grid. Left: Covariance fit showing the simulated covariance (red crosses) overlapped with the reconstructed ones (blue line). Right: Corresponding profiles, for the simulated profile (red line) and the reconstructed one (blue stars). Top: Reconstruction from a perfect covariance curve (black line in Figure 10). Bottom: Reconstruction from a noisy covariance curve (red crosses in Figure 10). The simulated and reconstructed profiles relative turbulence strengths are given in Table 2. For each graph we provide the value of the relative error between the simulated and reconstructed data.

Overall, when running a set of 100 noisy data simulations, the mean relative error on the profile reconstruction is 14% for the full range of altitude, 25% for the 5 to 24 km range and 5% for altitudes below 5 km.

## 5 FIRST MEASUREMENTS AND RESULTS AT THE SUTHERLAND SITE

The PML was deployed at the SAAO Sutherland observatory in South Africa during August 2011. During the PML observing campaign we also had a MASS-DIMM and a GSM running alongside it. On all nights that the PML was oper-ational, we were able to compare the seeing measurement to the values derived from the GSM and DIMM instruments, and the atmospheric profile to the measurements of the FA from MASS.

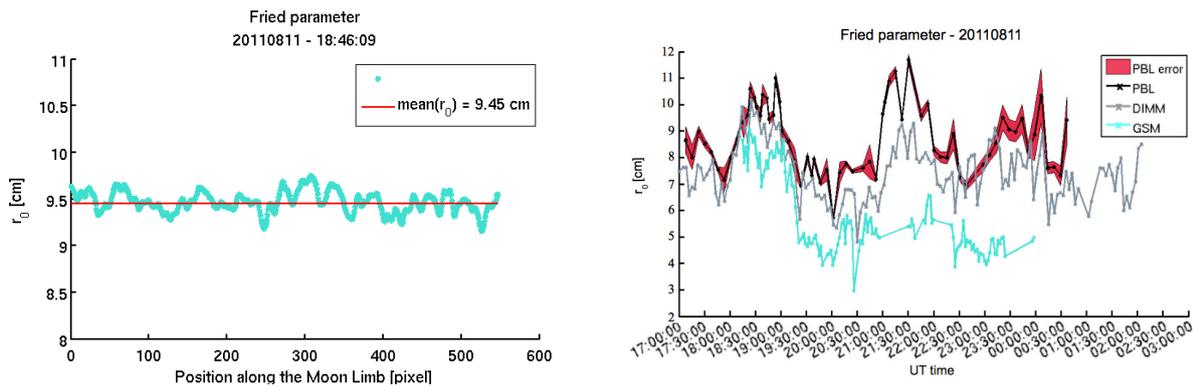

**Figure 12.** Fried parameter measurement. Left: Measurements of the Fried parameter for all positions along the Lunar limb on the night of 2011 August 11 at 19:46:09 UT with an average value of 9.45 cm. Right: Instrument comparison of the measured seeing on the night of 2011 August 11 with PML (red), GSM (green) and DIMM (grey).

### 5.1 Fried parameter measurements

We obtain the equivalent of a DIMM measurement for each field angle along the Lunar limb. The corresponding $r_0$ value for the acquisition is taken as the mean of all values along the limb. An example of Fried parameter measurement is shown in Figure 12. On the left hand-side one can see the measured Fried parameter for all positions along the Lunar limb for the acquisition obtained at 18:46:09 UT on the night of 2011 August 11. As given in the legend, the mean value of the Fried parameter for this acquisition was 9.45cm. On the right hand-side we have the seeing measurements through out the night along with the corresponding DIMM and GSM measurements. From the figure it is apparent that seeing measurement trends agree very well between PML and DIMM, but that the PML measures a better seeing. There was a height difference between the DIMM and PML setups. The DIMM entrance aperture is located approximately 1.5 m from the ground, while that of the PML one was positioned approximately 3 m from the ground. Hence we expect the PML to measure a higher $r_0$ value. In addition, even when fully opened, the sliding roof of the MASS-DIMM enclosure can still cause surface turbulence to worsen the seeing as seen from the instrument. On the other hand, one would expect the GSM and PML measurements to agree. There are 3 factors that could have contributed to the discrepancy between PML and GSM measurements. Even though both GSM and PML sits on a 1.5m pier, the GSM entrance aperture is slightly lower than that of the PML due to the instruments' respective sizes. In addition, they are not pointing at the same object and, hence, are not sensing the exact same part of the atmosphere. More importantly, and probably the main error contribution, there were contrast issues with the GSM during the campaign due to cirrus clouds, humidity as well as frost forming on the sub-apertures of the GSM unit.

### 5.2 Turbulence Profiles

The measured (red circles) and fitted (blue line) covariances for the acquisition obtained at 19:01:11 UT on the night of 2011 August 11 are shown in Figure 13 (left). The fitted covariance function corresponds to the best fit turbulence profile for that measurement. The retrieved profile is shown on the right hand-side of Figure 13. In order to verify both our profile reconstruction and seeing measurement with PML, we compared the seeing value obtained in the DIMM-mode ($r_0^{DIMM}$), as presented in section 5.1, and the value calculated from the full integration of the profile ($r_0^{profile}$). For the night of 2011 August 11 at 19:01:11 UT $r_0^{DIMM} = 9.04$ cm, as compared to $r_0^{profile} = 8.99$ cm. In addition, we compared the PML results with those from the MASS-DIMM instrument. We calculated $r_0$ from the integrated MASS profile which gives a value of 25.7 cm. Integrating the corresponding top layers of the PML profile (0.5 to 25 km) we obtain a $r_0$ of 27.1 cm. Both instruments agree well on the amount of turbulence located in the FA. However these are comparisons on a single acquisition. In order to obtain a sense of how well the data reconstruction performs, the value of the relative error between measured and reconstructed covariance for 125 measurements over two nights is displayed in the bottom graph of Figure 13. The mean relative error value is found to be 0.16%. Further verification is obtained by instrument cross comparison with MASS results on an entire night's data set.

Figure 14 shows the turbulence profile evolution throughout the night on 2011 August 11 from both PML and MASS measurements. The top figure shows the full PML profile with 33 layers. The corresponding MASS profile for the same night is displayed in the bottom left figure. On the bottom right we give the MASS weighting function, which defines the reconstruction range of the 6 altitudes. The dotted grey lines indicates the weighted central altitudes of each MASS bin. Note that those altitudes are just indicative. One can see from the triangular weighting functions of MASS that the contribution in each fixed layer could, in reality, be due to turbulence at higher and/or lower altitudes. Also, all the turbulence measured by the PML in the GL is unseen by the MASS. It is also worth noting that the scale of the turbulence strength on the PML and MASS profiles are not the same. This is due to a different dynamic range. Since the MASS uses less layers for the profiles reconstruction, it allocates more turbulence in each layers than does the PML reconstruction. The turbulence strength within individual MASS layers ranges from $5e^{-15}$ to $2e^{-12}$, while it ranges from $1e^{-16}$ to $5e^{-12}$ for the PML.

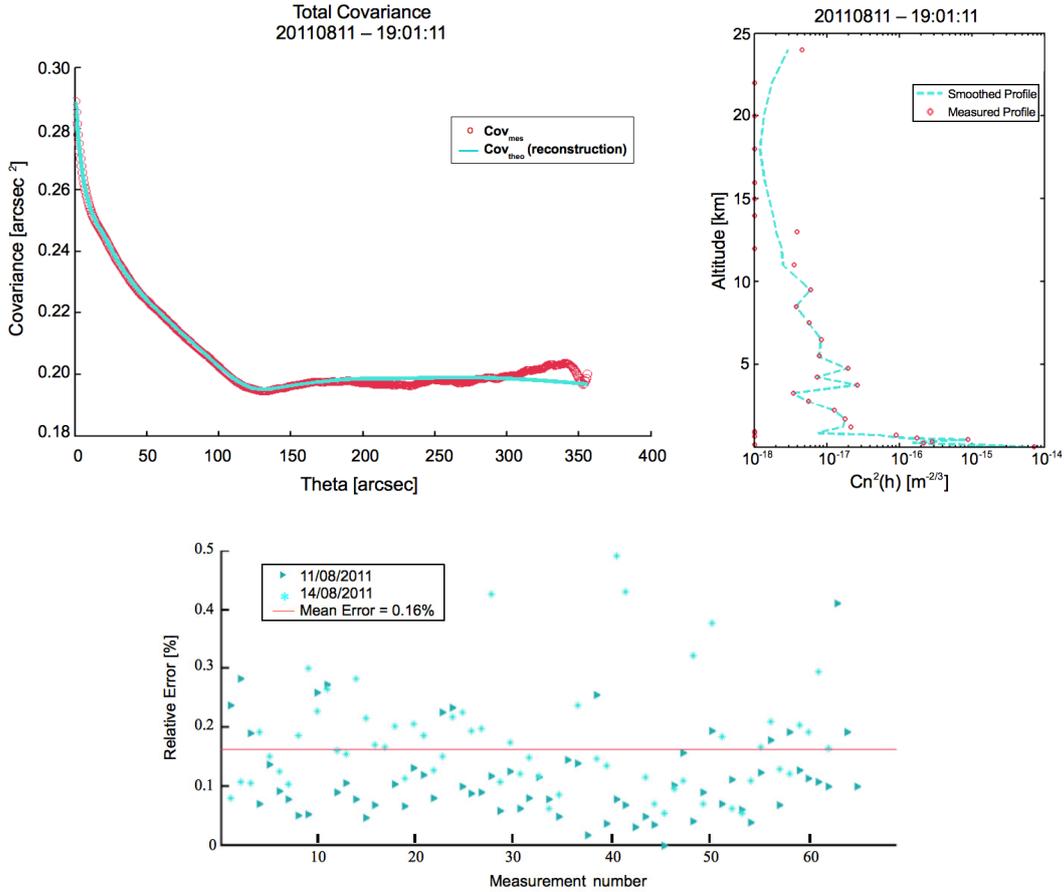

**Figure 13.** PML turbulence profile results. Top Left: measured covariance (red circles) and the best fit of the theoretical covariance (blue line) on 2011 August 11 at 19:01:11 UT. Top Right: retrieved PML turbulence profile from the best fit covariance. Bottom: Relative error between measured covariances and fitted ones for 125 measurements over 2 nights. The mean error value is 0.16%.

Comparing the FA turbulence from the PML (above the plain grey line) to the MASS profile, one can see a fairly good agreement between the two. We see strong turbulence around 500 m, and from 2 to 8 km before 18:00 UT, fading away later on with turbulence remaining mainly in the 2 and 8 km layers of MASS until 21:00 UT. These are seen around 1.7 km and between 4 to 10 km on the PML profile. After 23:00 UT, the 16 km layer of MASS becomes dominant, in particular around 00:00 UT where a peak of turbulence is also seen in the 4 and 8 km layers. Those are seen by PML from 4.5 to 18 km.

The PML profile has a clear advantage over the MASS one. Not only PML delivers a much higher altitude-resolution in the free atmosphere, but it also resolves the ground layer turbulence below 500 m , with a resolution of 100 m, unseen by the MASS instrument.

## 6 DISCUSSION AND CONCLUSIONS

The PML method proposed in this paper uses differential measurements, making it insensitive to tracking errors or telescope wind shake. Moreover, the use of the continuous Lunar limb provides a large range of separation angles, as compared to the double star used for SLODAR, allowing for the high-resolution of the altitude profile of the turbulence. The large number of separation angles available also permits a characterization of both the ground layer and free atmosphere where most other instruments are tuned to determine one or the other. The method was validated by testing it on both simulated synthetic data and cross comparison with MASS-DIMM and GSM results. Simulations showed that the reconstruction, using a simulated annealing method, was accurate within 14% of the real value, with higher error for altitudes above 5 km, mainly due to incorrect redistribution between adjacent layers. Further optimization of the simulated annealing inversion could potentially help to lower the error. Other reconstruction algorithms have been investigated, in particular that used in Blary et al. (2014) that led to lower errors.

The comparison to DIMM measurements, for the $r_0$ value, and MASS, for the $C_n^2(h)$ profile, shows good agreement in both cases. The great advantage of the PML over the MASS is its much higher altitude-resolution, with 33 layers spread through both the GL, with a vertical resolution of 100 m, and the FA, with resolution ranging from 500 m to 2 km, while the MASS has only 6 layers, limited to the FA alone.

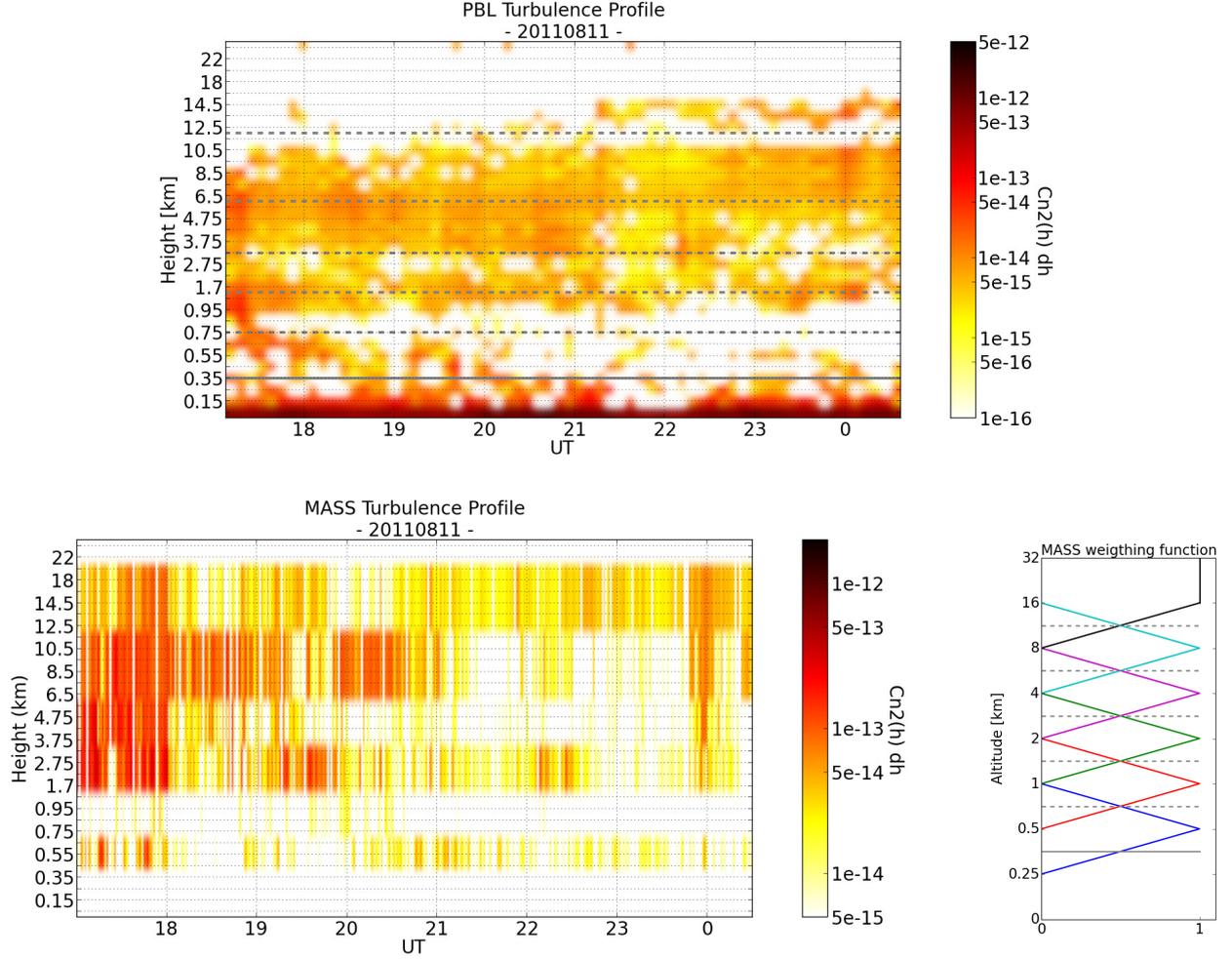

**Figure 14.** Turbulence profiles for the night of 2011 August 11. Top: PML measurements. Bottom left: MASS profile for the same night. Bottom right: MASS weighting functions. As an indication we reported the average separation between MASS layers, represented as dotted grey lines, on the PML profile (Top figure). In addition all PML layers below the plain grey line are unsensed by the MASS. The white bottom area on the MASS corresponds to this range of altitudes where there is no measurement. Note that the scale of PML and MASS profile are not the same as their data do not have the same dynamic range due to the different number of layers.

A first measurement of $\mathcal{L}_0(h)$ profile gave promising results, but further improvements, in particular to increase the number of layers in the reconstruction, are required.

In principle, both $\theta_0$ and $\tau_0$ can be retrieved from PML data. This is something that could be implemented in further data analysis. Also, more work is currently being done to speed up the data processing in order to have an automated system that can produce real-time measurements.

## 7 ACKNOWLEDGEMENTS

This work is based upon research supported by the National Research Foundation (NRF) via the multiwavelength grant and support to L.C., S.C and D.B.. Additional support was provided by the bilateral agreement in science and technology between France and South Africa, PROTEA.

Data analysis was performed on Matlab and graphs and plotting were produced with both matlab and matplotlib on Python.

In addition the authors would like to thank the staff of the South African Astronomical Observatory (SAAO).

This paper has been typeset from a T<sub>E</sub>X/L<sup>A</sup>T<sub>E</sub>X file prepared by the author.